Optimised information gathering in smartphone users

Arko Ghosh*, Jean-Pascal Pfister and Matthew Cook

Institute of Neuroinformatics
University of Zurich and ETH Zurich
Winterthurerstr. 190
CH-8057, Zurich
Switzerland

* Corresponding author

Arko Ghosh
Institute of Neuroinformatics
University of Zurich and ETH Zurich
Winterthurerstr. 190
CH-8057, Zurich
Switzerland

arko@ini.uzh.ch

Conflicts of interest:

Arko Ghosh is an inventor of the patent-pending technology used to track touchscreen interactions in this study. Arko Ghosh and Jean-Pascal Pfister are co-founders of QuantActions GmbH, a company focused on quantifying human behaviour through smartphone interactions.

Funding:

This study was funded by the Society in Science the Branco Weiss Fellowship and a research grant from the Holcim Stiftung, both awarded to author AG. MC and JPP were supported by the Swiss National Science Foundation.

Significance Statement:

The patterns underlying animal behaviour and their significance for the behavioural outcomes has remained unclear. Here we report that the temporal structure of human behaviour is distinct when sharing information vs. when checking for information on the smartphone. Playing back the discovered patterns in computer simulations we propose that humans introspect in the near term to maximise the proportion of successful checking attempts. Furthermore, the temporal pattern of sharing information was in tune with the pattern people typically used to check their phones. Studying a ubiquitous modern behaviour may help understand the gathering and sharing of resources in biological systems in general.

Abstract:

Human activities from hunting to emailing are performed in a fractal-like scale invariant pattern. These patterns are considered efficient for hunting or foraging, but are they efficient for gathering information? Here we link the scale invariant pattern of inter-touch intervals on the smartphone to optimal strategies for information gathering. We recorded touchscreen touches in 65 individuals for a month and categorized the activity into checking for information vs. sharing content. For both categories, the inter-touch intervals were well described by power-law fits spanning 5 orders of magnitude, from 1s to several hours. The power-law exponent typically found for checking was 1.5 and for generating it was 1.3. Next, by using computer simulations we addressed whether the checking pattern was efficient – in terms of minimizing futile attempts yielding no new information. We find that the best performing power law exponent depends on the duration of the assessment and the exponent of 1.5 was the most efficient in the short-term i.e. in the few minutes range. Finally, we addressed whether how people generated and shared content was in tune with the checking pattern. We assumed that the unchecked posts must be minimised for maximal efficiency and according to our analysis the most efficient temporal pattern to share content was the exponent of 1.3 – which was also the pattern displayed by the smartphone users. The behavioural organization for content generation is different from content consumption across time scales. We propose that this difference is a signature of optimal behaviour and the short-term assessments used in modern human actions.

Introduction

Surface mail correspondences, website visits and library loans, are apparently different from each other but according to quantitative explorations they all occur in bursts separated by long gaps(1-3). This pattern occurs across the time scales of examination and goes against the conventional assumption that the timing of human actions can be described using Poisson processes(4). Instead, the patterns are well approximated by the heavy-tailed power-law distribution. Apart from the power-law distributions in the timing of human actions, they are also apparent when insects, birds and mammals travel in search of food or abode(5, 6). For instance, lost bees attempt to find their way back home with short bursts of exploration separated by longer distance flights(7). The straight-line flight distances are distributed according to a power-law.

In theory, the way animals and birds search (the levy-flight pattern) is optimal when the food is scattered in a patchy environment(8). This pattern also increases the success rate when searching for a hidden or sparse target. This optimal strategy may be preserved in humans as the distances travelled by hunters and gatherers show the same scale-invariant search statistics(1). In contrast to the pattern of distances, it is unclear what the timing of the human actions is optimised for. Previous research focused on the generative process underlying the heavy-tailed power law distribution of action timings(4, 9, 10). According to one prominent theory the power-law distribution of inter-event times in email and mail correspondences is an emergent property of the priority-based decision process(4). In brief, this theory suggests that humans perform according to a task list where the task with the highest priority is executed first as opposed to being chosen randomly or according to the order in which they were added to the list. This theory can explain why the exact shape of the power-law distribution differs with the type of activity. For example, surface mail correspondences are captured by a power-law exponent of 1.5 whereas emails are captured by 1.0 (the lower the value the more the proportion of longer gaps). The current theoretical framework does not provide insights into the consequence of these behavioural patterns. Essentially, what would be the consequence of generating emails with an exponent of 1.5 instead of 1.0?

As checking online news and email correspondences follow the same power-law exponent (1.0), this data counter-intuitively suggests that the timing of information consumption is no different from generating and sharing it (2-4). A string of neuroscientific and psychological explorations indicate that the human brain is specialised for language and social interactions(11, 12). This distinct line of research raises the possibility that the behavioural pattern underlying information sharing is fundamentally different from the consumption. Smartphone behaviour offers a fresh avenue for exploring whether the previous findings generalise when the behaviour is unconstrained and when the primary interface necessary to share or consume information is ubiquitously present.

In this study, we examined a range of activities accomplished using the smartphone, from social networking to checking the weather. One purpose of this report was to explore the consequences of the smartphone behavioural patterns specifically for information gathering, in contrast to previous explorations that focused instead on the underlying generative process (for surface mails, short messages and emails)(2, 4, 9). Towards this goal, the broad range of activities on the phone was separated into two categories: (i) Information or content consumption – as in checking the weather and the messages posted by others and (ii) Information or content generation aimed towards sharing – as in composing 'tweets' for Twitter. We find that these two categories are in fact fundamentally different across time-scales and the exact power-law exponent impacts the efficiency of information gathering.

Results and discussion

We measured the timing of touchscreen interactions on the smartphone. To isolate the activity associated with information gathering we analysed interactions predominantly used towards activities such as checking the time, weather, message notifications and online searches (**Fig. 1**). The inter-touch intervals associated with such information gathering were well approximated for each person by power-law exponent, $\alpha = 1.5$ (median)[inter-quartile range, $iqr^{\alpha} = 0.16$, $n_{people} = 65$, $n_{taps}$ $1.4 \times 10^4$ (median)] for $\tau > 10^{3.0}$ ms (median), referred to as $\tau_{min}$ in the rest of the text, $iqr^{\tau min}= 10^{0.2}$ ms, (**Fig. 1 c,e**)]. Next, we quantified the keyboard use on social Apps such as twitter or Facebook which are used to generate and share content. The inter-touch intervals for the generation were well approximated for each person by $\alpha = 1.3$ (median), [$iqr^{\alpha}= 0.18$, $n_{people} = 65$, $n_{taps}$ $1.3 \times 10^3$ (median), $\tau_{min} = 10^{3.4}$ ms (median), $iqr^{\tau min} = 10^{0.3}$ ms, (**Fig. 1 d,f**)]. Not only were the power-law exponents largely consistent across the population, the power-law hypothesis could not be ruled out in 65% of the individual fits ($p > 0.1$, bootstrapped statistics comparing the empirical data to synthetic data drawn from the power-law distribution). The distribution of exponents for checking information was distinct from the exponents for content generation ($p = 2.8 \times 10^{-6}$, Wilcoxon signed rank test). By pooling across the population, we again recovered $\alpha = 1.5$ for checking information (for $\tau_{min} = 10^3$ ms) and $\alpha = 1.3$ for content generation (for $\tau_{min} = 10^4$ ms, for both type of activities $p > 0.1$, bootstrapped statistics and with $n_{taps} = 1.3 \times 10^6$ and $n_{taps} = 1.6 \times 10^5$ inter-events respectively) (**Fig. 1 c,d**). In sum, checking information on the phone involves a distinct pattern compared to generating content and the difference was captured by the power-law exponents of 1.5 and 1.3 respectively.

How does the power-law exponent observed for checking information on the phone compare with the other exponents in terms of efficient information gathering? To address this, we simulated with a wide range of power-law exponents (between 1 and 3) and measured the resulting efficiency for information gathering (**Fig. 2**). In our simulation we assumed that information is emitted by uncorrelated Poisson sources and this resulted in an inter-event distribution of external information with $\alpha = 2.0$ (13) (**Fig. 2 a,b**). The smartphone is typically carried by the users at all times but the actions on the phone are metabolically and cognitively expensive while actions that yield information are rewarding (14).

Formally, let us consider an assessment interval *[0,T]* (**Fig. 2 a**). This interval (period *T*) defines the period over which touchscreen actions and their consequences are integrated to determine the success rate (defined below). For example, an *impatient* system that cannot afford to be without information for too long (or with only short-term memory) would be set to a short T.

In this interval, external information is emitted at times $t_j^e$, j = 1,...,$n_e$ where the inter-emission times τ = $t_j^e$ - $t_{j-1}^e$ are distributed according to a power-law distribution with α = 2.0 (13) (**Fig. 2 a,b**). Checks are performed at times $t_j^c$, j = 1,...,$n_c$. A check at time $t_j^c$ is successful if there is at least one external event in the interval $[t_{j-1}^c, t_j^c]$. Formally, the success of a check at time $t_j{}^c$ is defined as

$$S^c(t_j^c) = \begin{cases} 1 & \text{if there is at least one event } t_i^e \in [t_{j-1}^c, t_j^c] \\ 0 & \text{else} \end{cases}$$

(Equation 1)

where it is assumed for notational convenience that $t_0^c = 0$. Over the $k^{th}$ assessment interval *[0,T]*, we can define the checking success rate *$SR^c(k)$* as

$$SR^c(k) = \begin{cases} \frac{1}{n_c}\sum_{j=1}^{n_c} S(t_j^c) & \text{if } n_c > 0 \\ 0 & \text{if } n_c = 0 \end{cases}$$

(Equation 2)

where *$n_c$* corresponds to the number of checks during the $k^{th}$ assessment interval. Since the individual checking times $t_j^c$ are random variables, the checking success rate *$SR^c(k)$* is also a random variable. We define the average checking success rate as

$$\overline{SR}^c = \frac{1}{n}\sum_{k=1}^{n} SR^c(k)$$

(Equation 3)

where *n* is the number of assessment periods over which the average checking success rate is calculated.

We explored the performance of checking patterns defined by the power-law exponents under different duration of assessments (*T*). The larger the duration of assessment the closer the best performing checking pattern got to *α* = 1.0 (**Fig. 2 c, d**). This draw towards *α* = 1.0 is intuitive as it is composed of longer gaps than the exponents of a higher value. With such long gaps, there is a high probability that new events are emitted within the gap and subsequently picked-up by the system. However, with the smaller durations of assessment both the low exponents near *α* = 1.0 and extremely high exponents are sub-optimal. At low exponents, there may be no checking attempt by the system at all in the assessment period leading to a checking success of 0 (see the *$n_c$ = 0* condition in Eq. 2), and at high exponents there may be too many failed checking attempts (see Eq. 1). $\alpha$ = 1.5, typically observed for the checking of information on the smartphone, is optimal at a $T$ of $10^{1.8}$ simulation units – one simulation unit was the minimum inter-check interval of 1 (**Fig. 2 d**). According to this framework, smartphone users frequently assess their behaviour in terms of information gathering.

Given the strong role of the duration of assessment ($T$) it must be further elaborated. Firstly, in the real world biological systems cannot wait endlessly for assessments for optimizing performance - making a parameter such as $T$ plausible. Secondly, it is also plausible that in biological systems the assessments occur continuously but the period from which the data points are gathered must be limited by memory or by the time constants associated with action control and information. Thirdly, how the $T$ of $10^{1.8}$ simulation units discovered above converts to time units is not clear. If we were to simply consider that the simulation unit corresponds to the $\tau_{min}$ estimated above (~1s), then humans assess their digital behaviour over a time period of about a minute. The temporal bounds of assessment may be a general feature of scale-invariant behavioural organization, as a recent study linked the optimal behaviour in spatial searches (akin to the levy flight) to cognitively defined temporal discounting (5).

People typically check the phone with $\alpha$ = 1.5, but does the exponent with which they generate and share content match the checking exponent? We assumed that the best exponent for sharable content generation maximizes the number of messages or posts that are seen by others before sending the next post or message (**Fig. 2 e**). We formalized this idea by first defining the message sending times $t_i^m$, $i = 1,...,n_m$, where $n_m$ is the number of messages sent during the interval $[0,T]$. We then defined the success of a message at time $t_i^m$ as

$$S^m(t_j^m) = \begin{cases} 1 & \text{if there is at least one event } t_j^c \in [t_i^m, t_{i+1}^m] \\ 0 & \text{else} \end{cases}$$

(Equation 4)

where it is assumed that $t_{n_m+1} = T$. Note that the definition of the success of a message looks into the future (i.e. it must be read in the future by someone) whereas the definition of the success of a check (see Eq. 1) looks into the past (it checks past information).

Similarly, to the checking case, we can define the message success rate as

$$SR^m(k) = \begin{cases} \frac{1}{n_m} \sum_{j=1}^{n_m} S(t_j^m) & \text{if } n_m > 0 \\ 0 & \text{if } n_m = 0 \end{cases}$$

(Equation 5)

where $n_m$ denotes the number of messages sent in the $k^{th}$ assessment period. The average message success rate is therefore defined as

$$\overline{SR}^m = \frac{1}{n}\sum_{k=1}^{n} SR^m(k)$$

(Equation 6)

As in the previous simulations for the optimal strategy for checking the phone, we searched for the best exponent for posting messages (**Fig. 2 f,g**). The previous set of analyses on the most efficient pattern to check the phone determined that humans introspect on their information gathering with an assessment period of $10^{1.8}$ simulation units. According to the simulations using Eq. 6, the optimal exponent to generate content at this period ($10^{1.8}$ simulation units) was $α = 1.3$ (**Fig. 2 h**). The same exponent was also observed in smartphone users when generating content. Our results suggest that for smartphone users the checking for information and the generation and sharing of information are optimised for the short-term and that these activities are optimised in relation to each other.

The behavioural structures found here are similar to what has been discovered for the inter-event times associated with gathering food in the sense that they display a heavy-tailed power-law distribution (15, 16). Specifically, the inter-event times associated with searching for food in the wild is well described by $α = 1.5$. In sea birds, this temporal pattern has been connected to the sparse distribution of food and by how the odour associated with the prey disperses to the bird (8, 15, 17). This theory associated with food odour does not easily map onto the ideas presented here on information gathering, optimal behaviour and the time windows of assessment. Still, our framework on information gathering may relate to the gathering of food. The objective functions explored here may be adapted to the gathering of food by simply substituting information with food. According to this framework, sea birds and fish may perform short-term evaluations of their behaviour when hunting – as short as the duration of assessment used by smartphone users when gathering information (18). For food, the duration of the assessment may be governed by metabolic needs (18). In spite of the apparent differences between gathering food and information, similar strategies may be employed to maximise the yield with as little effort as possible. $α = 1.3$ observed here for content generation is not typical in the temporal patterns expressed by the rest of biology (7, 15, 18). Perhaps this reflects the distinct behavioural and cognitive strategies used for sharing vs. gathering resources.

This study is complementary to the important set of explorations focused on establishing the presence of the heavy-tailed behavioural pattern in humans and animals and the corresponding processes that generate them (4, 7, 15, 19). This study characterizes the heavy-tailed behavioural pattern but focuses on the putative consequences on information gathering. The exponents of 1.5 and 1.3 are consistent with the priority-based generative process (2, 3, 9). Our work raises the possibility that putative consequences feeds back onto processes such as the priority-based decision process that govern behaviour across time scales.

Conclusion

Actions on the smartphone are patterned corresponding to the type of activity. We discovered two patterns of activity – one corresponding to information checking and another corresponding to content generation and sharing on the phone. The patterns were distinct across time scales with fewer long gaps for the checking compared to the generation. The checking behaviour was approximated with a heavy tailed power-law exponent of 1.5 and the generation was approximated with an exponent of 1.3. Our analysis using computer simulations suggests that the exponent of 1.5 is the most efficient for checking the phone when the system is assessed in the short-term (~ 1 min) and for the same short-term assessment the exponent of 1.3 is the most efficient for generating sharable content. In spite of the apparent difference between the two types of activity, the processes that optimise them may be deeply interlinked. Finally, the basic behavioural structure employed in smartphone use to gather information may also be employed in the rest of biology to gather resources.

Methods

Volunteers

Sixty-five volunteers were recruited through campus wide announcements at the University of Zurich and ETH-Zurich. The volunteers were between 20 and 45 years of age (30 females). The volunteers signed an informed consent and all of the observations were approved by the Kanton of Zurich enforcing the Swiss Human Experimentation Act. All the recruited users confirmed the use of only one smartphone during the study period and that it was not shared with anyone else.

Smartphone measurements

The side button presses used to turn-on the touchscreen and the touchscreen taps were recorded at a 5 ms resolution using a background App through the entire period of study (now available from QuantActions GmbH, Lausanne, Switzerland)(20). The keyboard operation was also logged by the App. The background recording occurred across all Apps and the App labels were additionally labelled and categorized as Social and Non-social categories. Social Apps were those that enabled users to share content with a circle of friends or acquaintances (some of the common social Apps were WhatsApp, Twitter, Facebook and Tinder).

Power-law fits and data analysis

The inter-event times over 1000 ms were processed as described in detail here (19). Briefly, the distribution of inter-event times is fitted to a power-law exponent using maximum likelihood estimates and the parameter $\tau_{min}$ was determined using the Kolmogorov-Smirnov (KS) distance. The empirical fit (KS distance) was then compared against bootstrapped synthetic power-law datasets (1000) with the null hypothesis that the empirical and synthetic datasets are drawn from the same distributions. The τ-min of the synthetic datasets was limited with a lower bound of 1000. All of the analysis was conducted by using Matlab 2015 (MathWorks, Natick, MA, USA) and run the scripts made available by Aaron Clauset (via tuvalu.santafe.edu/powerlaws/). The Wilcoxon ranking tests to compare the activities were performed using the statistical tool box on Matlab 2015.

Simulations

To evaluate the efficiency of power-law exponents for information gathering and content generation we constructed two separate simulations using Matlab with the objective functions described the main text. The objective functions were assessed for 1000 times, and power-law exponents between 1.1 and 3 were tested with a step size of 0.01. It is from these n = 1000 assessments that the best power-law exponent was determined. Each assessment block (of the n = 1000 assessments) were repeated 100 times to document the output variation of the simulation.

Figure Legends

Figure 1: The heavy-tailed distribution of inter-event times in smartphone users when engaged in distinct activities. (a) The activity on the smartphone was categorised as checking for information when it involved the swipes and taps, but not the pop-up keyboard use and not the *in-App* interactions on Social Apps (as in Twitter, WhatsApp or Facebook). The checking of notifications pushed by the Social or non-Social Apps were included in this category. (b) The activity was categorised as generating content – to share with others – when it involved the use of the keyboard within Social Apps. (c-d) The distribution of inter-event times in one user. Insert: pooled inter-event times from the sampled populations. The power-law exponent $\alpha$ well captures all intervals greater than $\tau_{min}$ marked with arrow. (e-f) The distribution of power-law exponents $\alpha$ in the sampled population. Insert: the corresponding distribution of $\tau_{min}$ values.

Figure 2: Computer simulations reveal the efficiency of different power-law exponents in checking and generating content. (a) Illustration of the simulations used to evaluate the best power-law exponents ($\alpha$) for checking information. (b) The external information emitted to the individual was abstracted by uncorrelated Poisson sources resulting in a heavy-tailed power-law distribution of inter-event times with $\alpha = 2$. (c) The efficiency of a range of $\alpha$ between 1.1 and 3 were evaluated while checking for the simulated external events – the exponents between 1.1 and 2 are plotted for illustration. (d) In the simulation, the best $\alpha$ for the checking was determined for various durations and $10^3$ assessments. The error bars represent the standard deviations of the corresponding mean values based on $10^2$ repetitions. Note that at the duration of assessment of $10^{1.8}$ time steps the exponent of 1.5 performed the best. Insert: the outputs of the $10^2$ repetitions used to determine the best performing $\alpha$ at the duration of assessment of $10^{1.8}$. (e) Illustration of the simulation used to assess the best performing $\alpha$ for content generation. (f) We explored the efficiency of a range of exponents between 1.1 and 3, and the exponents between 1.1 and 2 are plotted for illustration. (g) In the simulation, the inter-event times generated with $\alpha = 1.5$ was used for checking the generated content (as observed for checking in smartphone users). (h) Given the estimated assessment period of $10^{1.8}$ simulation units – as determined in 'd' – the best performing exponent was 1.3 for generating content.

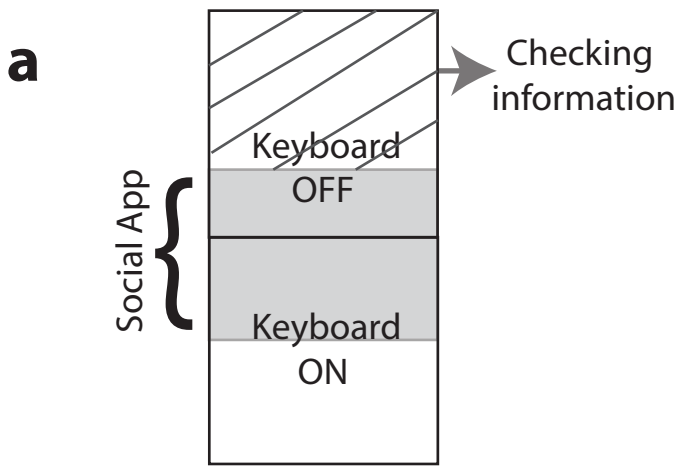 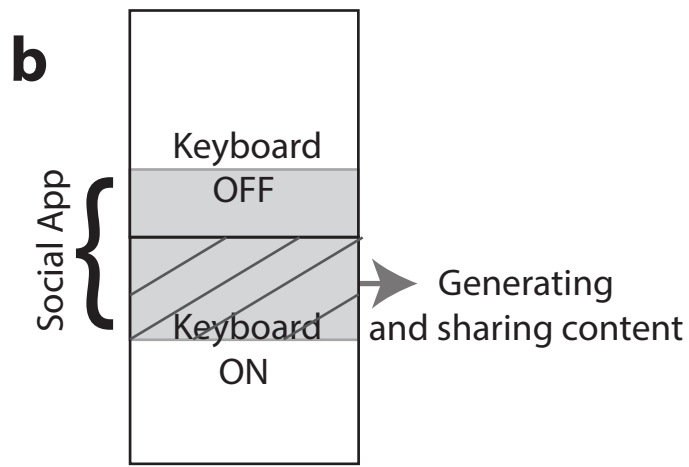

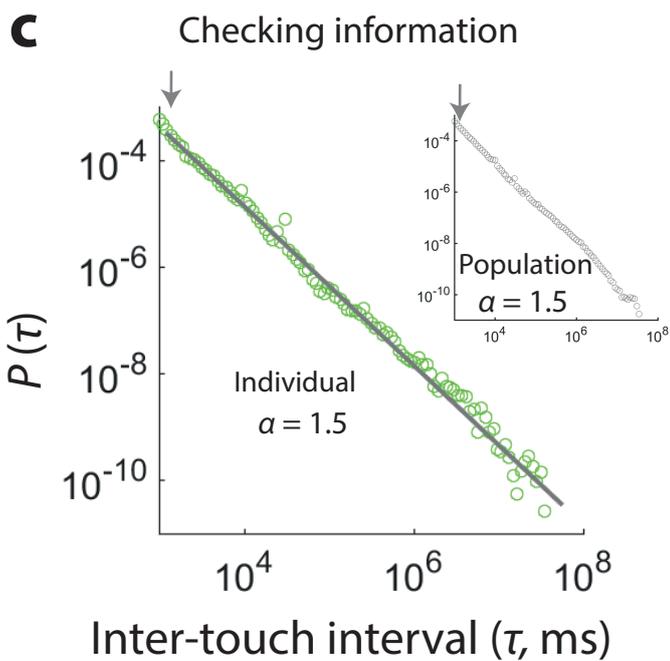 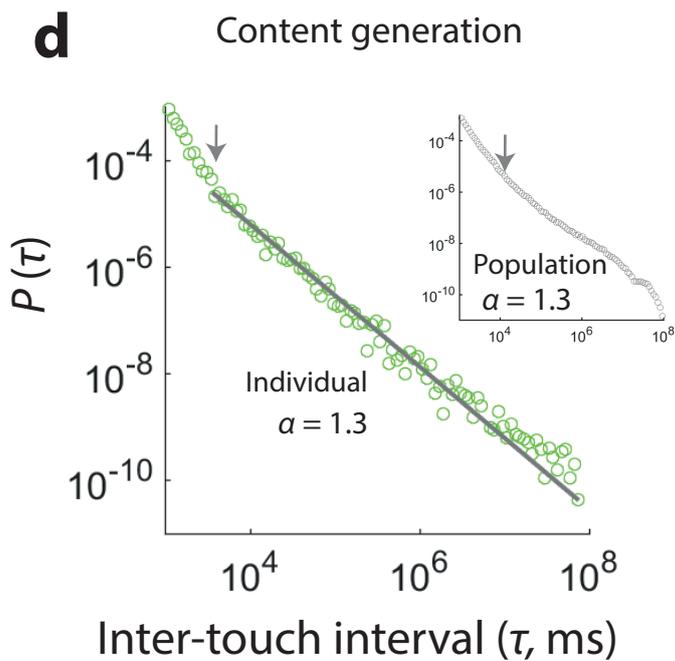

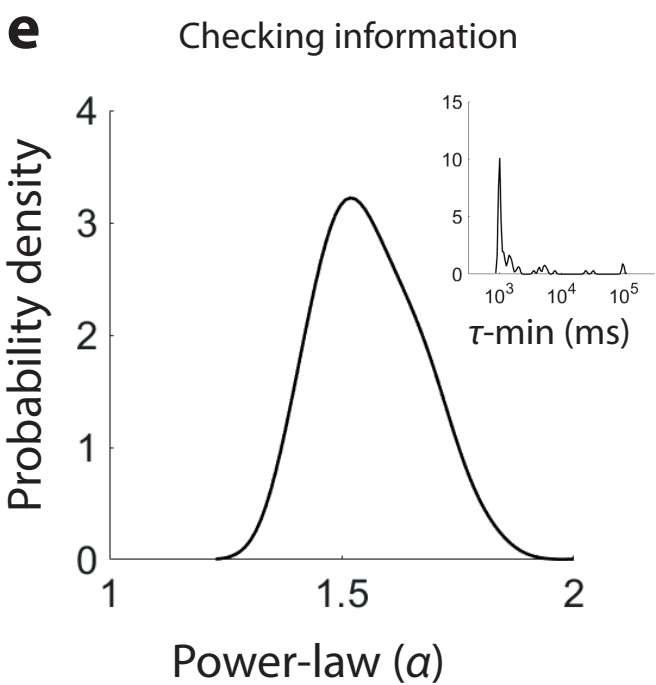 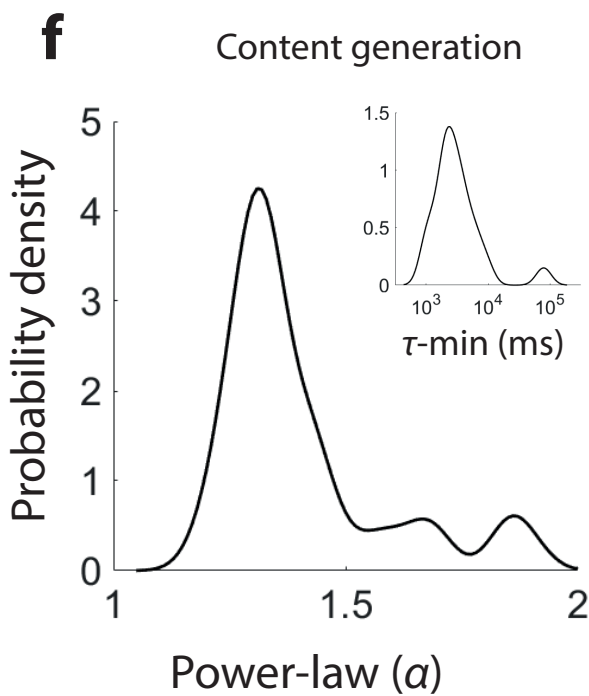

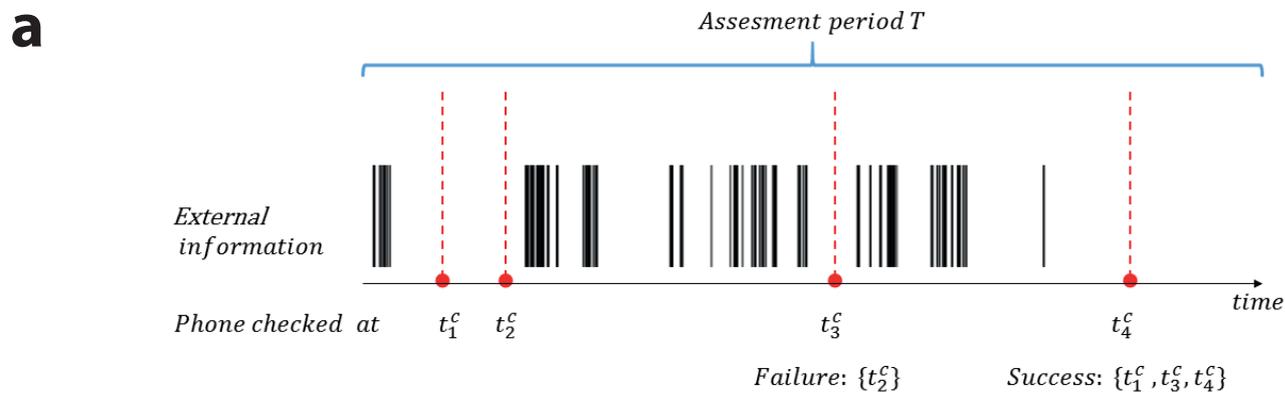

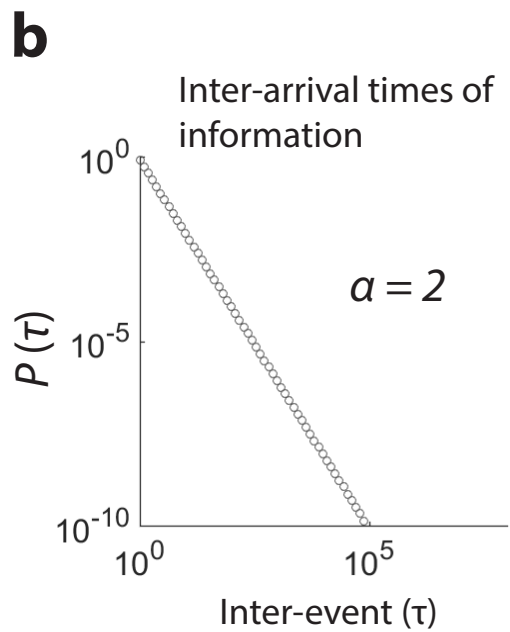

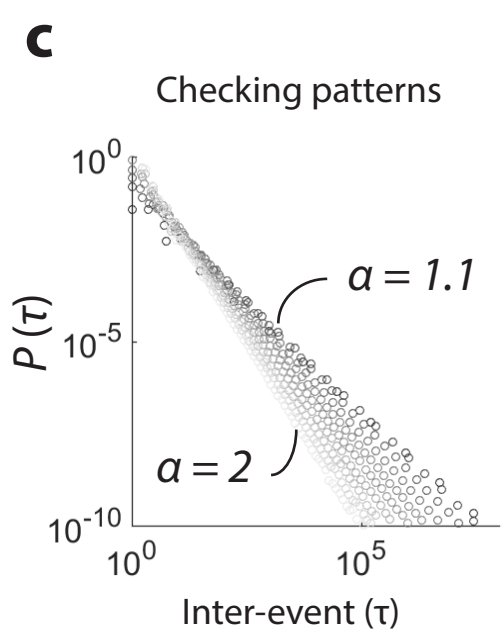

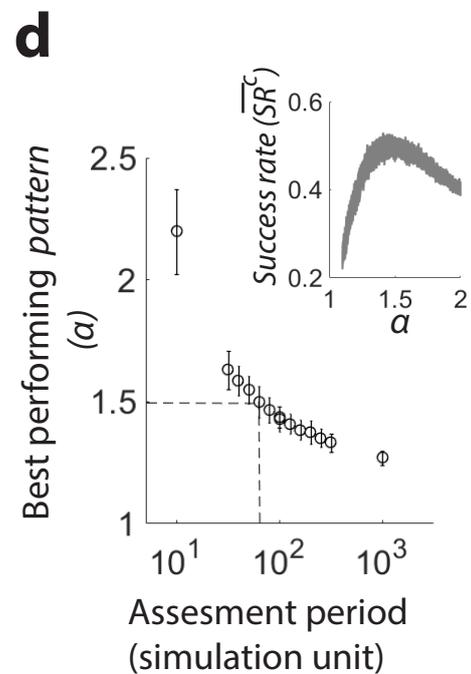

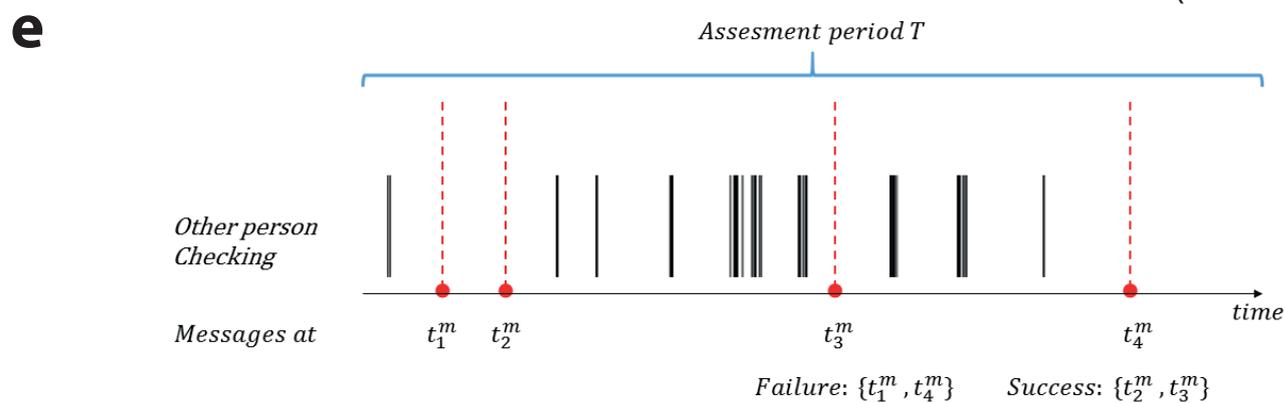

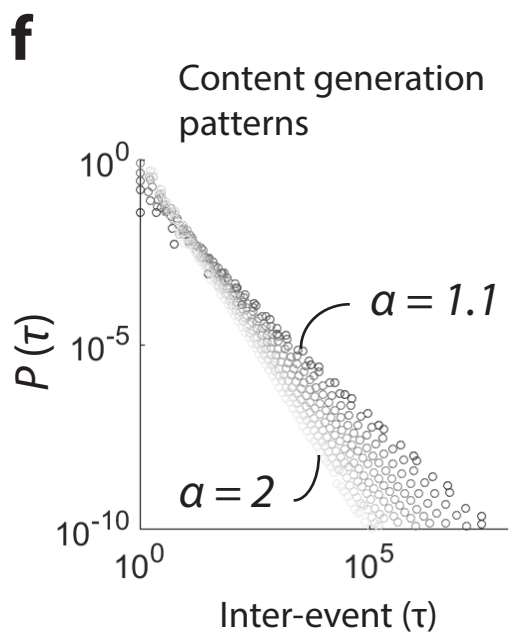

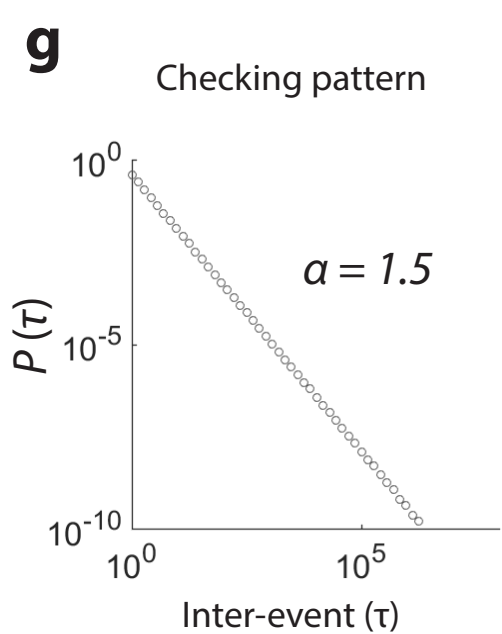

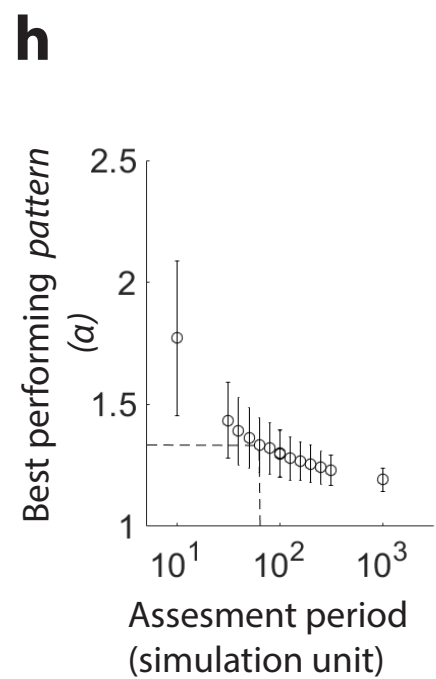